\documentclass[aps,prl,twocolumn,superscriptaddress,longbibliography]{revtex4-2}
\usepackage{graphicx}
\usepackage{amssymb,amsmath}
\usepackage{bm}
\usepackage{dcolumn}
\usepackage{float}
\usepackage[OT1]{fontenc} 
\usepackage{url}
\usepackage{mathrsfs}
\usepackage{slashed,comment}
\usepackage{color}
\usepackage{verbatim}
\usepackage{enumitem}
\usepackage{soul,physics}
\usepackage[driverfallback=dvipdfm]{hyperref}
\hypersetup{pdfpagemode=FullScreen,colorlinks=true,breaklinks,urlcolor=blue,linkcolor=blue,citecolor=blue}

\usepackage{subfigure}
\usepackage{amssymb}
\usepackage{bm}
\usepackage{graphicx}
\usepackage{amsmath,amssymb}

\begin{document}
  
  \title{Characterizing Fermionic Non-Gaussianity in the Sachdev-Ye-Kitaev Model \\ via Replica Twist Entropy}

  \author{Ning Sun}
  \affiliation{State Key Laboratory of Surface Physics \& Department of Physics, Fudan University, Shanghai, 200438, China}

  \author{Pengfei Zhang}
  \thanks{PengfeiZhang.physics@gmail.com}
  \affiliation{State Key Laboratory of Surface Physics \& Department of Physics, Fudan University, Shanghai, 200438, China}
  \affiliation{Hefei National Laboratory, Hefei 230088, China}

  \date{\today}

  \begin{abstract}
  Understanding the non-equilibrium dynamics of quantum many-body systems plays a central role in modern many-body physics. Intriguing insights emerge from investigating how quantum resources, which are essential for achieving unambiguous quantumness, are generated and spread under chaotic quantum dynamics. In fermionic systems, non-Gaussianity constitutes an important resource beyond entanglement, characterizing the deviation of generic quantum states from free-fermion states. In this Letter, we propose using the \textit{replica twist entropy}, defined through the expectation value of a replica rotation operator with a tunable angle $\alpha$, as a natural probe of non-Gaussianity in pure fermionic states. After deriving its behavior for typical many-body states, we establish a general framework for analyzing the replica twist entropy in solvable Sachdev-Ye-Kitaev models in the large-$N$ limit. Applying this framework to the dynamics of the thermofield double state, we uncover a novel dynamical phase diagram in the parameter space of $\alpha$ and evolution time $t$, exhibiting spontaneous $Z_2$ symmetry breaking at $\alpha=\pi/4$ and universal cusps as a function of $\alpha$ in the long-time regime. Our results pave the way for studying non-Gaussianity in strongly correlated fermionic systems and reveal a new class of dynamical transitions probed by non-Gaussianity.
  \end{abstract}
    
  \maketitle

  \emph{ \color{blue}Introduction.--} Strong interactions between fermions not only renormalize their masses and velocities but also give rise to a rich variety of quantum phases, including fractional quantum Hall states~\cite{RevModPhys.71.S298} and non-Fermi liquids~\cite{RevModPhys.94.035004}. Alongside traditional approaches based on analytical approximations and numerical algorithms, quantum information theory offers a complementary perspective by characterizing interaction effects in terms of fermionic non-Gaussianity, also known as fermionic magic, which serves as a quantum resource for achieving intrinsic quantumness in many-body systems~\cite{Hebenstreit:2019dly,Cudby:2023pdf,Reardon-Smith:2024efn,Hakkaku:2021wpp,Gigena:2020jih}. States with vanishing non-Gaussianity admit an efficient classical representation within the fermionic Gaussian-state formalism~\cite{Terhal:2001aml,Bravyi:2004bud}. Non-Gaussianity therefore characterizes the deviation of a quantum state from the set of fermionic Gaussian states. This parallels the role of magic resource, or non-stabilizerness, in qubit systems, where the Gottesman--Knill theorem guarantees efficient classical simulation of Clifford circuits acting on stabilizer states~\cite{Gottesman:1997zz,Gottesman:1998hu,PhysRevA.57.127,PhysRevA.70.052328}.

  Several concrete measures of fermionic non-Gaussianity have been proposed from different perspectives~\cite{Gottlieb:2005zuj,Gottlieb:2006xhw,Pachos:2022vjc,Bittel:2024qma,Lumia:2023ofv,Lyu:2024yaq,Coffman:2025usz,Bellomia:2025bwg,Sierant:2025fax,Tarabunga:2026ath,Leone:2026wac,Turkeshi:2026bvs}, enabling studies of the universal behavior of interaction-generated quantum resources in many-body systems~\cite{Santra:2025dsm,Ares:2026jes,Ares:2026vbb,Garcia-Garcia:2026fpf,Matsuda:2026ttw,Falcao:2026eew}. For example, non-Gaussianity can be characterized by examining violations of Wick's theorem~\cite{Gottlieb:2006xhw,Gottlieb:2005zuj,Pachos:2022vjc}. Related measures based on the correlation matrix include fermionic antiflatness~\cite{Sierant:2025fax}. However, measures based solely on the correlation matrix cannot distinguish states with identical two-point functions but differen,t higher-order correlations, limiting their ability to characterize the complex correlation structures in chaotic fermionic systems. Alternative measures, such as the magic R\'enyi entropy, employ fermionic quantum convolution~\cite{Lyu:2024yaq,Coffman:2025usz} to probe correlations of arbitrarily high order. Their evaluation typically involves observables acting on three or more copies of the density operator, posing additional challenges for ites numerical simulations in many-body systems.

  In this Letter, we introduce the \textit{replica twist entropy}, a one-parameter family of measures that characterizes fermionic non-Gaussianity in pure states from a symmetry perspective. These measures are defined through the expectation value of a replica rotation operator with a tunable angle $\alpha$ and probe complex correlations using only two copies of the density operator. After characterizing its behavior for typical many-body states, we develop a general framework for evaluating the replica twist entropy in Sachdev-Ye-Kitaev (SYK) models, a celebrated class of solvable models of quantum chaos~\cite{Sachdev:1992fk,kitaev2014talk,maldacena2016remarks}. Focusing on the thermofield double (TFD) state~\cite{Israel:1976ur,Maldacena:2001kr}, of broad interest in condensed matter physics~\cite{Liu:2024mme,Weinstein:2024fug}, quantum information~\cite{Gao:2016bin,Maldacena:2017axo,Susskind:2017nto,Gao:2018yzk,Brown:2019hmk,Gao:2019nyj,Schuster:2021uvg,Jafferis:2022crx}, and gravitational physics~\cite{Maldacena:2001kr}, we uncover a rich dynamical phase diagram as a function of $\alpha$ and the evolution time $t$, with features reminiscent of Ising universality. These include spontaneous $Z_2$ symmetry breaking at $\alpha=\pi/4$ and universal cusps as a function of $\alpha$ at long times. Our results establish the replica twist entropy as a diagnostic of non-Gaussianity that can reveal dynamical critical behavior hidden from simple observables.

  \emph{ \color{blue}Replica Twist Entropy.--} For generality, we consider fermionic systems consisting of $2N$ Majorana fermions $\chi_j$, with $j\in\{1,2,\ldots,2N\}$. The Majorana operators satisfy the canonical anticommutation relations $\{\chi_j,\chi_k\}=2\delta_{jk}$. We focus on quantifying the deviation of a pure quantum state with fixed fermion parity from the set of Gaussian states. To motivate the definition of the replica twist entropy, we recall that any pure Gaussian state $|\phi\rangle$ can be obtained from a reference Gaussian state $|0\rangle$ in the same parity sector through a unitary generated by a quadratic Hamiltonian: $|\phi\rangle=U_G(\phi)|0\rangle$, with $U_G(\phi)=e^{\frac{1}{2}\sum_{j<k}h_{jk}(\phi)\chi_j\chi_k}$. Introducing multiple replicas reveals a symmetry that distinguishes Gaussian evolution from generic unitary dynamics. Writing $U_G(\phi)^{\otimes 2}=e^{\frac{1}{2}\sum_{r}\sum_{j<k}h_{jk}(\phi)\chi_j^{(r)}\chi_k^{(r)}}$, with replica indices $r\in\{1,2\}$, we see that this operator is invariant under arbitrary $SO(2)$ rotations in replica space, $\chi_j^{(a)}\mapsto\sum_b R^{ab}(\alpha)\chi_j^{(b)}$~\cite{Cai2007OnTT,Sierant:2026hcd,Braccia:2026dwx,Lastres:2026nbx}. These rotations are implemented by the replica rotation operator:
  \begin{equation}
  U_R(\alpha)=\exp\Bigg(\frac{\alpha}{2}\sum_{j=1}^{2N}\chi_j^{(1)}\chi_j^{(2)}\Bigg):= \exp(i \alpha \Lambda).
  \end{equation}
  Consequently, the replicated Gaussian state $|\phi\rangle^{\otimes 2}$ is invariant under replica rotations $U_R(\alpha)|\phi\rangle^{\otimes 2}=|\phi\rangle^{\otimes 2}$. In the presence of interactions, the replicated state generally loses this rotational symmetry while remaining invariant under replica permutations with $\alpha=\pi/2$. This distinction plays a dominant role in unverisal dynamics of free-fermion systems, governing a range of phenomena including measurement-induced quantum phases~\cite{Zhang:2021klq,Bao:2021bhc,PhysRevLett.127.140601} and deep thermalization~\cite{Lucas:2022rkh,Bejan:2024emk,Sun:2026ahc}.

  Motivated by this symmetry analysis, we define the \textit{replica twist entropy} of a generic fermion state $\ket{\psi}$ using the logarithm of the expectation value of the replica rotation operator:
  \begin{equation}
   \mathcal{P}(\alpha,\psi):=-\ln\!\left[\langle\psi|^{\otimes 2}U_R(\alpha)|\psi\rangle^{\otimes 2}\right], \ \ \ \alpha\in(0,\pi/2).
  \end{equation}
  It satisfies the basic requirements for a non-Gaussianity measure on pure states: (i) Faithfulness: it is real and nonnegative, and vanishes if and only if $|\psi\rangle$ is Gaussian. (ii) Gaussian invariance: it remains invariant under arbitrary Gaussian unitaries. (iii) Additivity: it is additive under tensor products, $\mathcal{P}(\alpha,\psi\otimes\phi)=\mathcal{P}(\alpha,\psi)+\mathcal{P}(\alpha,\phi)$. The nonnegativity in property (i) is discussed in detail below, while the remaining properties follow directly from the definition and the symmetry considerations above. Expanding the replica twist entropy at small $\alpha$ leads to 
  \begin{equation}
  \mathcal{P}(\alpha,\psi)\approx \frac{\alpha^2}{2}Q(\psi),\ \ \ \ \ Q(\psi)=\langle\psi|^{\otimes 2}\Lambda^2|\psi\rangle^{\otimes 2}.
  \end{equation}
  Here, $Q(\psi)$ is the non-Gaussianity measure studied in Ref.~\cite{Coffman:2025usz} and is proportional to the lowest-order fermionic antiflatness~\cite{Sierant:2025fax}. These measures depend only on two-point correlation functions, whereas the replica twist entropy at finite $\alpha$ captures correlations of arbitrarily high order in the many-body state $\ket{\psi}$. Generalizing the replica twist operator to more than two replicas is also straightforward. 

  The replica twist entropy can be measured experimentally through two complementary protocols, detailed in the Supplementary Material~\cite{SM}. First, the replica twist overlap can be reconstructed using randomized Gaussian measurements~\cite{Zhao2021FermionicShadows} on repeated preparations of a single copy. Second, preparing two identical copies allows the overlap to be extracted from the full counting statistics of complex fermions formed by pairing Majoranas across the replicas. In both protocols, the angle dependence is obtained by classical postprocessing of the same measurement records, allowing the replica twist entropy to be evaluated as a function of $\alpha$ without performing a separate set of experiments for each angle.

  \emph{ \color{blue}General Properties.--} To derive relations satisfied by the replica twist entropy, we expand the unitary operator in the Pauli-string basis, denoting the basis elements by $\chi_S$ for concieness:
  \begin{equation}\label{eq:expansion}
  e^{-\mathcal{P}}=\sum_{S}\cos\left(\frac{\alpha}{2}\right)^{2N-|S|} \sin\left(\frac{\alpha}{2}\right)^{|S|} \left|\langle \psi|\chi_S|\psi\rangle\right|^2.
  \end{equation}
  Here, $|S|$ denotes the size of the Majorana string $\chi_S$, defined as the number of Majorana operators it contains. Since $\ket{\psi}$ has fixed fermion parity, only terms with even $|S|$ contribute to the sum. Consequently, $\mathcal{P}(\alpha,\psi)=\mathcal{P}(-\alpha,\psi)$ upon extending the domain of $\alpha$ to include negative values. Combining this symmetry with the identity $U_R(\frac{\pi}{2})|\psi\rangle^{\otimes 2}=|\psi\rangle^{\otimes 2}$ for any fermionic state $\ket{\psi}$ with fixed parity, we find
  \begin{equation}\label{eq:reflection}
  \mathcal{P}(\alpha,\psi)=\mathcal{P}(-\alpha,\psi)=\mathcal{P}({\pi}/{2}-\alpha,\psi).
  \end{equation}
  This reflection defines a $Z_2$ transformation that leaves the point $\alpha=\pi/4$ invariant. The associated symmetry will play an important role in our subsequent discussion of the growth of non-Gaussianity.

  Since $U_R(\alpha)$ is unitary, we have $|\langle\psi|^{\otimes 2}U_R(\alpha)|\psi\rangle^{\otimes 2}|\leq 1$. Together with the reality of $\mathcal{P}(\alpha,\psi)$ established by the expansion \eqref{eq:expansion}, this implies the lower bound $\mathcal{P}(\alpha,\psi)\geq 0$. Isolating the contributions from the identity and fermion parity operators yields the upper bound $\mathcal{P}(\alpha,\psi)\leq \mathcal{P}_m(\alpha)$, where
  \begin{equation}
  \mathcal{P}_m(\alpha)=-\ln\bigg[\cos\left(\frac{\alpha}{2}\right)^{2N}+\sin\left(\frac{\alpha}{2}\right)^{2N}\bigg].
  \end{equation}
  Applying the reflection transform \eqref{eq:reflection} yields the complementary bound $\mathcal{P}(\alpha,\psi)\leq \mathcal{P}_m(\pi/2-\alpha)$. In the thermodynamic limit $N\rightarrow \infty$, combining these bounds gives
  \begin{equation}\label{eq:bound}
  \frac{\mathcal{P}(\alpha,\psi)}{N}\leq \begin{cases}
  -2\ln \cos\left(\frac{\alpha}{2}\right)& \alpha\in(0,\frac{\pi}{4}]\\
   -2\ln \cos\left(\frac{\pi}{4}-\frac{\alpha}{2}\right)& \alpha\in(\frac{\pi}{4},\frac{\pi}{2})
   \end{cases}.
  \end{equation}
  Interestingly, the bound exhibits a cusp at $\alpha=\pi/4$: its slope is positive at $\alpha=\pi/4^-$, as determined by $-\ln \cos\left(\frac{\alpha}{2}\right)$, and abruptly reverses sign at $\alpha=\pi/4^+$ due to the reflection symmetry. 

  This intriguing nonanalyticity in $\alpha$ also appears in chaotic quantum states $\ket{\psi}$, which we model as Haar-random states within a fixed parity sector. We compute the Haar average of $e^{-\mathcal{P}}$ using the moment formula~\cite{Yoshida:2017non}. The derivation is provided in the Supplementary Material~\cite{SM}, and the result reads
  \begin{equation}
  e^{-\mathcal{P}_{\text{Haar}}}=\frac{2^N}{2^N+2}\sum_{r=0}^3\cos\left(\frac{\alpha}{2}+\frac{r\pi}{4}\right)^{2N}.
  \end{equation}
  For $\alpha\neq \pi/4$, a single term dominates the sum as $N\rightarrow \infty$, and the result saturates the bound \eqref{eq:bound} up to exponentially small corrections, reproducing its cusp in the thermodynamic limit. At $\alpha=\pi/4$, two dominant contributions become degenerate, yielding an additional correction of $-\ln 2$ to $\mathcal{P}_{\text{Haar}}$. This behavior is closely analogous to the Page curve for the entanglement entropy of pure states~\cite{Page:1993wv,Page:2013dx}, with the angle $\alpha$ playing the role of the subsystem size.

  \emph{ \color{blue}The SYK Model.--} Having established the general properties of the replica twist entropy, we now illustrate how it reveals novel dynamics of non-Gaussianity in the solvable SYK model~\cite{Sachdev:1992fk,kitaev2014talk,maldacena2016remarks}. The SYK model describes Majorana fermions with random all-to-all interactions and admits a systematic large-$N$ expansion. We consider the SYK Hamiltonian acting on the first $N$ Majorana modes as:
\begin{equation}\label{eqn:H}
H=\sum_{1\leq j<k<l<m\leq N}\frac{J_{jklm}}{4}\chi_{j}\chi_{k}\chi_{l}\chi_{m}.
\end{equation}
The numerical factors ensure consistency with the normalization conventions in Ref.~\cite{maldacena2016remarks}. The random couplings $J_{jklm}$ are independent Gaussian variables with 
  \begin{equation}
  \overline{J_{jklm}}=0\ \ \ \ \ \ \overline{J_{jklm}^2}=3!J^2/N^3.
  \end{equation}
  The SYK model provides a tractable setting for studying quantum resources beyond entanglement in many-body systems~\cite{Zhang:2022yaw}. Previous studies have examined non-stabilizerness through numerical simulations at finite $N$~\cite{Bera:2025pfp,Jasser:2025myz,njgn-fksh,Malvimat:2026oqf} and large-$N$ expansions~\cite{Zhang:2025rky,Sun:2026ewq,Bettaque:2026vpl}, revealing a dynamical transition in quantum magic. More recently, fermionic antiflatness in the SYK model was evaluated using disorder-averaged two-point functions and found to evolve smoothly in time~\cite{Garcia-Garcia:2026fpf}.
 
  We consider TFD states~\cite{Israel:1976ur,Maldacena:2001kr}, a class of pure states that admit a convenient path-integral representation and have attracted broad interest~\cite{Liu:2024mme,Weinstein:2024fug,Gao:2016bin,Maldacena:2017axo,Susskind:2017nto,Gao:2018yzk,Brown:2019hmk,Gao:2019nyj,Schuster:2021uvg,Jafferis:2022crx,Maldacena:2001kr}. To construct the TFD state, we first define a maximally entangled state $\ket{\text{EPR}}$ between the first $N$ Majorana modes and the remaining $N$ modes by imposing $(\chi_j+i\chi_{j+N})\ket{\text{EPR}}=0$ for all $j=1,\cdots,N$~\footnote{For later convenience, we leave the EPR state unnormalized and choose $\langle \text{EPR}|\text{EPR}\rangle=2^{N/2}$.}. The TFD state at inverse temperature $\beta$ and time $t$ is then defined as
  \begin{equation}
  |\text{TFD}(t)\rangle=|\text{TFD}'(t)\rangle/\sqrt{Z_\beta}=e^{-(\frac{\beta}{2}+it)H}\lvert\mathrm{EPR}\rangle/\sqrt{Z_\beta},
  \end{equation}
  Here, $Z_\beta=\text{tr}_N[e^{-\beta H}]$ is the thermal partition function for the first $N$ Majorana modes and ensures the normalization of the TFD state. We then evaluate the disorder average of the replica twist entropy as
  \begin{equation}
  \mathcal{P}(\alpha,t)=-\overline{\ln\!\left[\langle\text{TFD}(t)|^{\otimes 2}U_R(\alpha)|\text{TFD}(t)\rangle^{\otimes 2}\right]}.
  \end{equation}  
  In general, this evaluation requires introducing disorder replicas~\cite{altland2010condensed}. In the SYK model, however, the replica-diagonal structure~\cite{Kitaev:2017awl} allows us to interchange the disorder average and the logarithm in the limit of $N\rightarrow\infty$. The same reasoning allows us to replace the partition function $Z_\beta$ with its disorder average. Our task therefore reduces to evaluating the following partition function
  \begin{equation}
  \mathcal{Z}(\alpha,t)=\overline{\langle\text{TFD}'(t)|^{\otimes 2}U_R(\alpha)|\text{TFD}'(t)\rangle^{\otimes 2}}.
  \end{equation}
  The replica twist entropy is then given by $\mathcal{P}(\alpha,t)=-\ln\mathcal{Z}(\alpha,t)+2\ln Z_\beta=-\ln\mathcal{Z}(\alpha,t)+\ln\mathcal{Z}(0,t)$.

  We calculate the partition function by generalizing the standard large-$N$ expansion used in SYK thermodynamics. The key observation is that the replica rotation operator $U_R(\alpha)$ is generated by a quadratic fermionic operator and therefore modifies only the free Green's function, preserving the large-$N$ structure. Leaving the details to the Supplementary Material~\cite{SM}, we write the saddle-point equations for the partition function $\mathcal{Z}(\alpha,t)$ as
  \begin{equation}
  \begin{aligned}
  G_{ss'}=[(G^{0,\alpha})^{-1}-\Sigma]^{-1}_{ss'}, \ \ \ \Sigma_{ss'}=J^2 f_sf_{s'}G_{ss'}^3,
  \end{aligned}
  \end{equation}
  where we have introduced the bilocal Green's function field $G_{ss'}=\frac{1}{2}\langle \chi_j(s)\chi_j({s'})\rangle$ and the associated self-energy field $\Sigma_{ss'}$, which is independent of the index $j\in\{1,\cdots, N\}$ due to permutation symmetry. The real parameter $s\in\{0,4t+2\beta\}$ labels the entire contour, and $f_s=1,i,-i$ for imaginary-time, forward real-time, and backward real-time evolution, respectively. The $\alpha$-dependence enters through the free Green's function $G^{0,\alpha}_{ss'}$, defined as the two-point function at $J=0$. For convenience, we split the full evolution contour into four branches $(k_1,b_1,k_2,b_2)$, each of length $t+\beta/2$, and use $u\in\{0,t+\beta/2\}$ to parametrize the evolution within each branch. These branches correspond to the ket and bra in replica 1 and the ket and bra in replica 2, respectively. The free Green's function explicitly reads
  \begin{equation}
  G^{0,\alpha}=G^0_{uu'} I\otimes I-i\frac{\cos \alpha}{2}\sigma_y\otimes I-i\frac{\sin \alpha}{2}\sigma_x\otimes \sigma_y.
  \end{equation}
  Here, we use the convention that, in $A\otimes B$, $A$ acts in the ket/bra space and $B$ acts in the replica space. We also define $G^0_{uu'}=\frac{1}{2}\text{sgn}(u-u')$. After obtaining the saddle-point solutions, we evaluate the partition function from the on-shell action as
  \begin{equation}
  -\frac{\ln \mathcal{Z}(\alpha,t)}{N}=\frac{1}{2}\ln \det[G]+\frac{3}{8}\int dsds'~G_{ss'}\Sigma_{ss'}.
  \end{equation}

  The $Z_2$ transformation discussed in the previous section also manifests in the path-integral representation. We implement this transformation through a combination of branch permutation and fermion parity:
\begin{equation}
(\chi_{k_1},\chi_{b_1},\chi_{k_2},\chi_{b_2})\rightarrow (\chi_{k_1},\chi_{b_2},-\chi_{k_2},\chi_{b_1}).
\end{equation}
A direct calculation shows that this transformation maps $G^{0,\alpha}$ to $G^{0,\pi/2-\alpha}$, thereby implementing the reflection discussed above\footnote{This transformation can be combined with replica permutation to form a larger group $D_4$.}. In particular, the system is invariant under this transformation at $\alpha=\pi/4$. If the symmetry is not spontaneously broken, the correlation between $k_1$ and $k_2$ vanishes. A nonzero correlation between these branches therefore signals spontaneous symmetry breaking, motivating the definition of the order parameter $\sigma=G_{\beta+2t^+,0^+}$. Deviations from $\alpha=\pi/4$ explicitly break the symmetry and play the role of an external magnetic field in the Ising model. This correspondence motivates an Ising-like description of the dynamical transition, as we now explicitly demonstrate.

  \begin{figure}[t]
    \centering
    \includegraphics[width=0.99\linewidth]{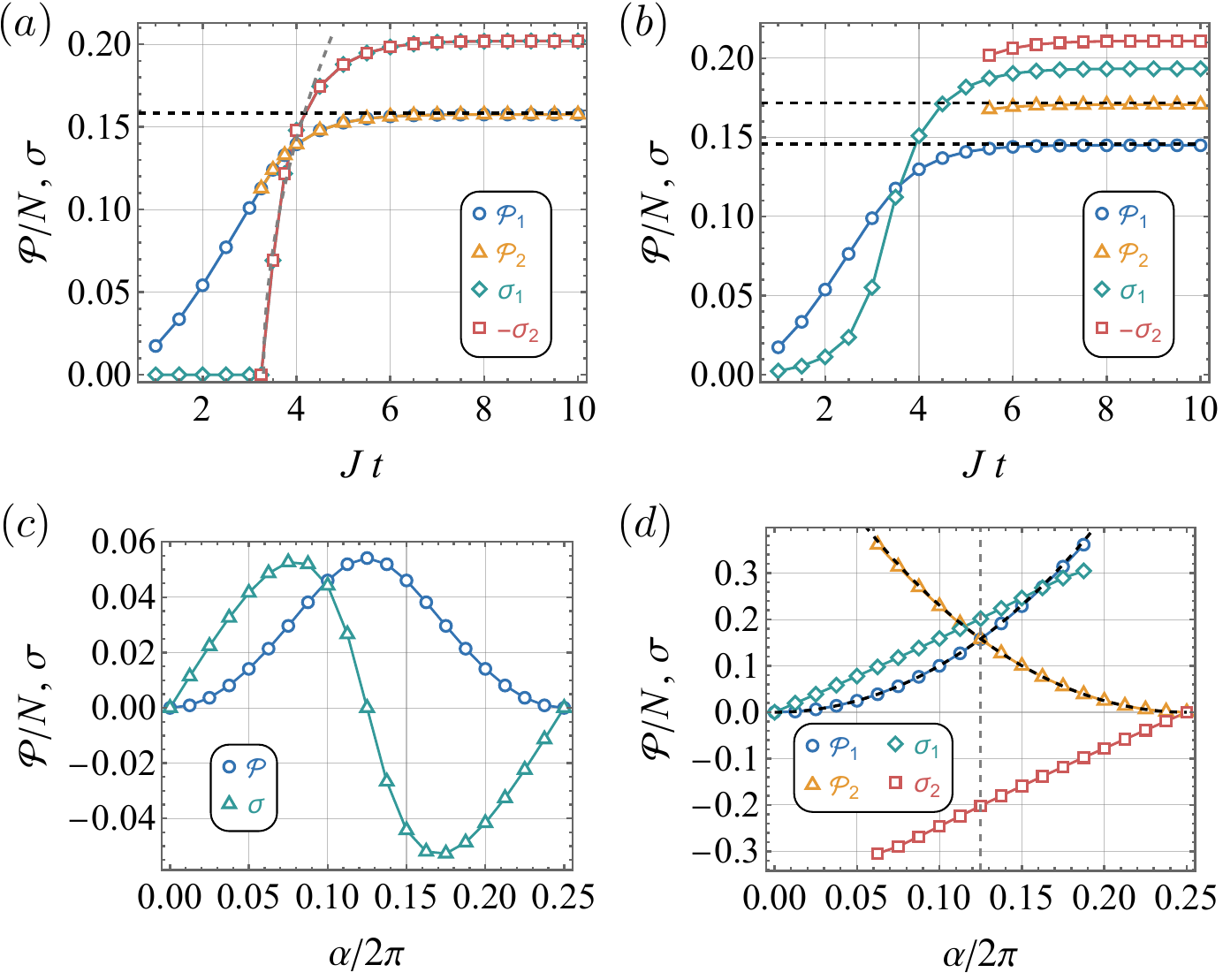}
    \caption{Numerical results for the replica twist entropy density $\mathcal{P}/N$ and the order parameter $\sigma$ in the SYK model at fixed $\beta J=1$. Panels (a-b) show results as a function of the evolution time $Jt$ for (a) $\alpha/2\pi=1/8$ and (b) $\alpha/2\pi=3/25$. The gray dashed line in panel (a) shows a fit to $c_0\sqrt{J(t-t_c)}$, with fitting parameters $(c_0,t_c)$. Panels (c-d) show results as a function of the rotation angle $\alpha$ for (c) $Jt=2$ and (d) $Jt=8$. In all panels, the black dashed lines show the prediction for Haar-random states, which saturates the bound \eqref{eq:bound}. }
    \label{fig:num}
  \end{figure}

  \emph{ \color{blue}Numerical Results.--} Obtaining an analytic solution of the saddle-point equations is challenging due to the absence of translation symmetry caused by the insertion of the replica rotation operator. We therefore analyze the corresponding equations numerically by discretizing the evolution contour and representing both the Green's function field and the self-energy field as matrices. Similar methods have been employed to calculate the R\'enyi entropy~\cite{Chen:2020wiq,Zhang:2020kia} and the stabilizer R\'enyi entropy~\cite{Sun:2026ewq}. For a given pair of physical parameters $Jt$ and $\alpha$, multiple saddle-point solutions may coexist. In this case, we add subscripts to both $\mathcal{P}$ and $\sigma$ to distinguish them. In the large-$N$ limit, the physical result is determined by the dominant saddle, which has the smallest on-shell action and therefore the smallest replica twist entropy.

  Results for both the replica twist entropy density $\mathcal{P}/N$ and the order parameter $\sigma$ are presented in FIG.~\ref{fig:num} at fixed $\beta J=1$. We first consider the reflection-symmetric angle $\alpha=\pi/4$ in FIG.~\ref{fig:num}(a). At early times, the order parameter vanishes while $\mathcal{P}/N$ grows. Near $Jt_c\simeq 3.3$, two saddle-point solutions emerge with equal actions, $\mathcal{P}_1=\mathcal{P}_2$, and opposite order parameters, $\sigma_1=-\sigma_2$, as required by the $Z_2$ symmetry. This onset is consistent with a second-order transition into a phase with spontaneously broken replica symmetry, further supported by a fit to $c_0\sqrt{J(t-t_c)}$ (shown in FIG.~\ref{fig:num}(a) by the gray dashed line). At later times, $\mathcal{P}/N$ approaches the Haar-random-state prediction. Entropy saturation is thus accompanied by a qualitative change in replica correlations. Moving away from $\alpha=\pi/4$ lifts the degeneracy between the two saddles. FIG.~\ref{fig:num}(b) shows results at $\alpha/2\pi=3/25$, where the dominant branch has $\sigma_1>0$ throughout the plotted time range. The transition is rounded into a crossover, consistent with the angle displacement acting as a symmetry-breaking field. A second solution with $\sigma_2<0$ appears at later times but has a larger action and therefore does not determine the physical entropy. At long times, $\mathcal{P}_{1/2}/N$ approaches the values associated with the leading and subleading bounds in \eqref{eq:bound}, respectively, indicated by the black dashed lines. The subdominant solution therefore introduces no additional transition along this time cut.

  \begin{figure}[t]
    \centering
    \includegraphics[width=0.8\linewidth]{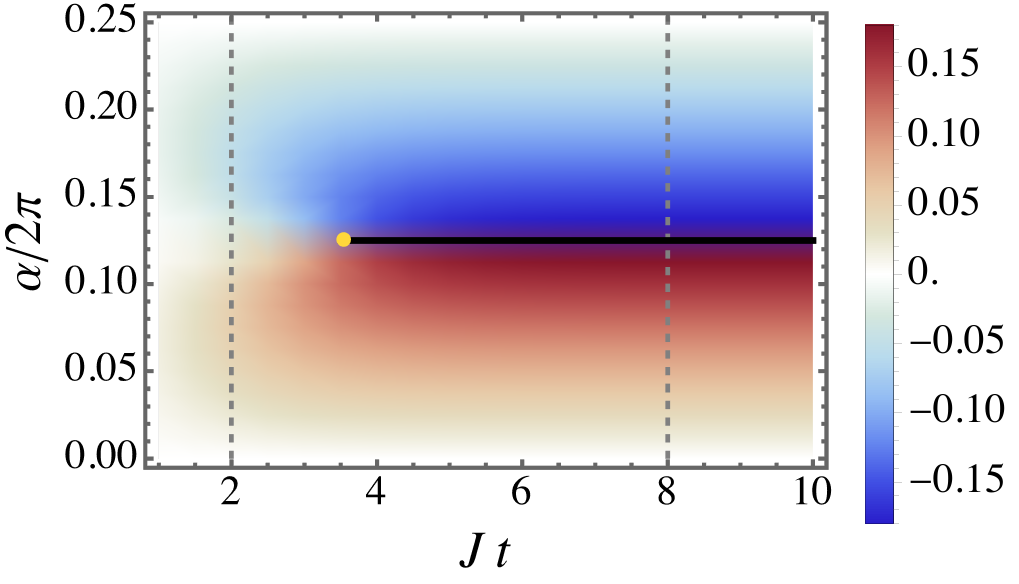}
    \caption{Dynamical phase diagram for the replica twist entropy in the SYK model at fixed $\beta J=1$, showing the order parameter $\sigma(\alpha,t)$. The black solid line marks the first-order transition, and the yellow point denotes the critical point. The gray dashed lines indicate the time slices shown in panels (c-d) of Fig.~\ref{fig:num}.}
    \label{fig:num2}
  \end{figure}

  Next, we examine the angle dependence at fixed time. At $Jt=2$, FIG.~\ref{fig:num}(c) shows a smooth entropy profile with a maximum at $\alpha=\pi/4$. The order parameter changes sign continuously and vanishes at this angle, consistent with an unbroken reflection symmetry. At $Jt=8$, the system exhibits spontaneous symmetry breaking at $\alpha=\pi/4$. Consequently, two saddle-point branches coexist near this angle, as shown in FIG.~\ref{fig:num}(d). They are exchanged under reflection, with $\mathcal{P}_1(\alpha,t)=\mathcal{P}_2(\pi/2-\alpha,t)$ and $\sigma_1(\alpha,t)=-\sigma_2(\pi/2-\alpha,t)$, and are therefore degenerate at the symmetric angle. As $\alpha$ passes through $\pi/4$, the dominant solution switches from branch 1 to branch 2. This switch produces a jump in the physical order parameter and a cusp in the continuous entropy, signaling a first-order transition. This behavior is analogous to tuning an external magnetic field through zero in the ferromagnetic phase of the Ising model. Putting these results together yields the dynamical phase diagram in FIG.~\ref{fig:num2}. The replica twist entropy thus reveals universal features of nonequilibrium dynamics captured by an Ising-like effective description. We emphasize that a finite twist angle $\alpha\approx \pi/4$ is essential for observing this nontrivial phase structure, which is absent in the small-$\alpha$ regime, consistent with previous studies of fermionic antiflatness~\cite{Garcia-Garcia:2026fpf}.

  \emph{ \color{blue}Discussions.--}
  In this Letter, we introduced the replica twist entropy as a two-copy diagnostic of fermionic non-Gaussianity in pure states that captures many-body correlations of arbitrarily high order. We established its general properties and determined its value for Haar-random states, providing a benchmark for chaotic quantum systems. We then developed a general framework for analyzing the replica twist entropy in the SYK model. Applying this framework to TFD states, we uncovered an Ising-like dynamical phase diagram as a function of the rotation angle $\alpha$ and the evolution time $t$. In particular, the growth of non-Gaussianity is accompanied by spontaneous symmetry breaking in replica correlations at $\alpha=\pi/4$. In the symmetry-broken phase at late times, the replica twist entropy develops a cusp as $\alpha$ is tuned through $\pi/4$. Our results establish the replica twist entropy as a diagnostic of fermionic non-Gaussianity in chaotic quantum systems, revealing information-theoretic phases that are invisible to simple observables.

  We conclude with several remarks. First, the cusp as a function of the rotation angle resembles those found in studies of disorder operators, where cusps serve as signatures of superfluidity~\cite{Wang:2023csw}, nontrivial topology~\cite{Mao:2025igj}, entanglement phase transitions~\cite{Tian-GangZhou:2023jsj}, and mixed-state anomalies~\cite{PhysRevLett.133.106503}. The replica twist entropy similarly involves the expectation value of a replica rotation operator analogous to a conventional disorder operator. Second, the present study focuses on the simplest setting, where symmetry breaking arises solely in replica space. Additional symmetries within each replica may lead to symmetry-enriched dynamical phases of non-Gaussianity. Finally, while the replica twist entropy characterizes non-Gaussianity in pure states, whether it or its generalizations can extract useful information about non-Gaussianity in mixed states remains an interesting open question. We leave these questions for future work.

\vspace{5pt}
  \textit{Acknowledgement.}
  This project is supported by the NSFC under grant 12604404 and 12374477, the Shanghai Rising-Star Program under grant number 24QA2700300, the Quantum Science and Technology-National Science and Technology Major Project 2024ZD0300101, the Xuemin Institute of Advanced Studies at Fudan University, and the Shanghai Qiguang Natural Science Development Foundation.

\vspace{5pt}
  \textit{Note Added.}
  While finalizing our study, we became aware of an independent work~\cite{Hoshino:2026fsk} investigating the dynamics of non-Gaussianity in the SYK model using the magic R\'enyi entropy. The authors also identified an analogous dynamical transition arising from the spontaneous breaking of replica permutation symmetry.

\bibliography{main.bbl}

\end{document}


\title{Supplementary Material: Characterizing Fermionic Non-Gaussianity in the Sachdev-Ye-Kitaev Model via Replica Twist Entropy}

\author{Ning Sun}
\affiliation{State Key Laboratory of Surface Physics and Department of
Physics, Fudan University, Shanghai 200438, China}

\author{Pengfei Zhang}
\email{PengfeiZhang.physics@gmail.com}
\affiliation{State Key Laboratory of Surface Physics and Department of
Physics, Fudan University, Shanghai 200438, China}
\affiliation{Hefei National Laboratory, Hefei 230088, China}

\date{\today}
  \begin{abstract}
In this Supplemental Material, we present (1) the replica twist entropy of Haar-random states, (2) the saddle-point equations for the SYK model, and (3) experimental protocols for measuring the replica twist entropy using randomized measurements and two-copy occupation measurements.
  \end{abstract}
    
  \maketitle

  \section{The replica twist entropy of Haar-random states}
  In this section, we detail the calculation of the replica twist entropy for Haar-random states in a fixed fermion parity sector, which we take to be even without loss of generality. For a system with $2N$ Majorana fermions, this subspace has dimension $d=2^{N-1}$. Using the second moment of Haar-random states~\cite{Yoshida:2017non}, we obtain the following identity for any Hermitian operator $O$:
  \begin{equation}
  \mathbb{E}_{\mathrm{H}}[\left|\langle \psi|O|\psi\rangle\right|^2]=\int d U~\langle 0|U^\dagger O U|0\rangle^2=\frac{\text{tr}_+(O^2)+\text{tr}_+(O)^2}{d(d+1)}
  \end{equation}
  Here, $dU$ denotes the Haar measure over unitaries $U$ acting within the even-parity subspace, and $\text{tr}_+$ denotes the trace over this subspace. To determine the replica twist entropy for typical states, we choose $O=\chi_S$, where $\chi_S$ is a Majorana string operator, yielding
  \begin{equation}
\mathbb{E}_{\mathrm{H}}[\langle\gamma_S\rangle^2]
=
\begin{cases}
1, & |S|=0 \text{ or } 2N, \\[6pt]
\dfrac{1}{d+1}, & 0<|S|<2N,\quad |S| \in\text{ even}.
\end{cases}
\end{equation}
Using this result, we evaluate the replica twist entropy through its expansion in Majorana strings, obtaining
 \begin{equation}\label{eq:expansion}
 \begin{aligned}
  e^{-\mathcal{P}_{\text{Haar}}}&\equiv \mathbb{E}_{\mathrm{H}}[e^{-\mathcal{P}}]=\sum_{|S| \in\text{ even}}\cos\left(\frac{\alpha}{2}\right)^{2N-|S|} \sin\left(\frac{\alpha}{2}\right)^{|S|}\mathbb{E}_{\mathrm{H}}[\langle \psi|\chi_S|\psi\rangle^2]\\
  &=\frac{d}{d+1}\left(\cos\left(\frac{\alpha}{2}\right)^{2N}+\sin\left(\frac{\alpha}{2}\right)^{2N}\right)+\frac{1}{d+1}\sum_{|S| \in\text{ even}}\cos\left(\frac{\alpha}{2}\right)^{2N-|S|} \sin\left(\frac{\alpha}{2}\right)^{|S|}\\
  &=\frac{d}{d+1}\left(\cos\left(\frac{\alpha}{2}\right)^{2N}+\sin\left(\frac{\alpha}{2}\right)^{2N}\right)+\frac{1}{2(d+1)}\sum_{\pm}\left(\cos\left(\frac{\alpha}{2}\right)\pm\sin\left(\frac{\alpha}{2}\right)\right)^{2N}\\
  &=\frac{2^N}{2^N+2}\sum_{r=0}^3\cos\left(\frac{\alpha}{2}+\frac{r\pi}{4}\right)^{2N}.
  \end{aligned}
  \end{equation}
  This is the result presented in the main text. In the limit of $N\rightarrow\infty$, we retain only the largest term in the sum and replace the overall prefactor by $1$, yielding
  \begin{equation}\label{eq:bound}
  {\mathcal{P}_{\text{Haar}}}=\begin{cases}
  -2N\ln \cos\left(\frac{\alpha}{2}\right)+O(e^{-N})& \alpha\in(0,\frac{\pi}{4})\\
   N\ln\frac{4}{2+\sqrt{2}}-\ln 2+O(e^{-N})& \alpha=\frac{\pi}{4}\\
   -2N\ln \cos\left(\frac{\pi}{4}-\frac{\alpha}{2}\right)+O(e^{-N})& \alpha\in(\frac{\pi}{4},\frac{\pi}{2})
   \end{cases}.
  \end{equation}
  Comparing this result with the upper bound presented in the main text, we find that Haar-random states asymptotically saturate the bound on the replica twist entropy.

  \begin{figure}[t]
    \centering
    \includegraphics[width=0.6\linewidth]{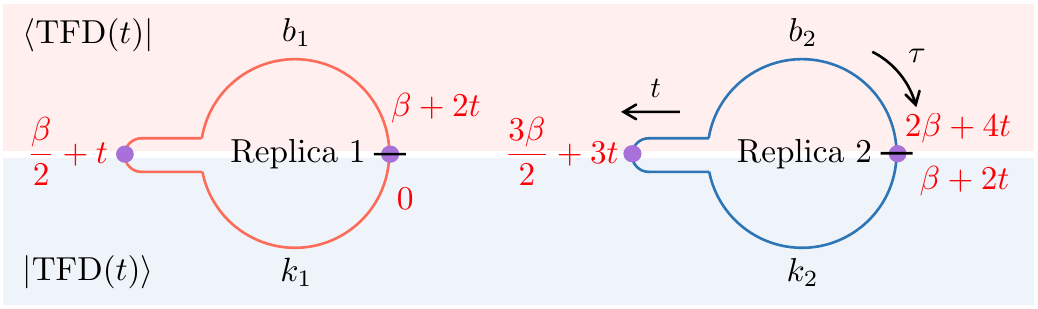}
    \caption{Illustration of the path-integral contour for the replica twist entropy, comprising two replicas with branches denoted by $(k_1,b_1,k_2,b_2)$. A single parameter $s\in(0,2\beta+4t)$ parametrizes the entire contour. The arcs represent imaginary-time evolution, while the horizontal lines represent real-time evolution. The purple dots mark the insertions of the replica rotation operators.  }
    \label{fig:contour}
  \end{figure}

  \section{The saddle-point equations for the SYK model}
  In this section, we detail the derivation of the saddle-point equations for the SYK model presented in the main text. Our aim is to express the partition function $\mathcal{Z}(\alpha,t)$ in the path-integral formalism. For completeness, we recall its definition:
  \begin{equation}
  \mathcal{Z}(\alpha,t)=\overline{\langle\text{TFD}'(t)|^{\otimes 2}U_R(\alpha)|\text{TFD}'(t)\rangle^{\otimes 2}},\ \ \ \ \ \ U_R(\alpha)=\exp\Bigg(\frac{\alpha}{2}\sum_{j=1}^{2N}\chi_j^{(1)}\chi_j^{(2)}\Bigg).
  \end{equation}
  By discretizing both real-time and imaginary-time evolution into infinitesimal steps and inserting resolutions of the identity in terms of fermionic coherent states, we express the partition function as a functional integral over Majorana fields. As discussed in the main text, we introduce a single real parameter $s\in (0,2\beta+4t)$ to label the path-integral contour. The convention for $s$ is illustrated by the red labels in FIG.~\ref{fig:contour}. The purple dots mark the insertions of the replica rotation operators, which introduce additional $\alpha$-dependent quadratic terms in the Hamiltonian. Putting all ingredients together, we find 
\begin{equation}\label{eq:act}
  \begin{aligned}
 &\mathcal{Z}(\alpha,t)=\int dJ_{jklm}P(J_{jklm})\int  \prod_{j=1}^ND\chi_{j}(s)~\exp(-S[\chi_j(s)]),\\
  &S=\int_0^{2\beta+4t} dsds'~\sum_{j}\frac{1}{4}\chi_{j}(s)(\delta_{ss'}\partial_{s'}-\sigma_{ss'}(\alpha)) \chi_{j}(s')+ \int_0^{2\beta+4t} ds~\sum_{j<k<l<m}\frac{J_{jklm}f_s}{4}\chi_{j}\chi_{k}\chi_{l}\chi_{m}.
  \end{aligned}
  \end{equation}
  Here, $P(J_{jklm})$ is the Gaussian distribution of the random couplings $J_{jklm}$, and $\sigma_{ss'}(\alpha)$ represents the contribution from $U_R(\alpha)$. Rather than specifying $\sigma_{ss'}(\alpha)$, it is more convenient to write the resulting free two-point function $G^{0,\alpha}_{ss'}$ at $J=0$. This function can be calculated directly from its definition in the operator representation. For example, when the two Majorana operators are inserted on the same branch of the contour $(k_1,b_1,k_2,b_2)$, we have 
  \begin{equation}
  G^{0,\alpha}_{ss'}\equiv \frac{1}{2}\langle \chi_j(s)\chi_j(s')\rangle=\frac{\text{sgn}(s-s')}{2}\frac{\langle\text{EPR}|^{\otimes 2}U_R(\alpha) \chi_j^{(1)}\chi_j^{(1)}|\text{EPR}\rangle^{\otimes 2}}{\langle\text{EPR}|^{\otimes 2}U_R(\alpha) |\text{EPR}\rangle^{\otimes 2}}=\frac{\text{sgn}(s-s')}{2}.
  \end{equation}
  When one Majorana operator $\chi_j(s)$ is inserted on $b_1$ and the other, $\chi_j(s')$, on $k_1$, we find
  \begin{equation}
  G^{0,\alpha}_{ss'}\equiv \frac{1}{2}\langle \chi_j(s)\chi_j(s')\rangle=\frac{1}{2}\frac{\langle\text{EPR}|^{\otimes 2}\chi_j^{(1)}U_R(\alpha) \chi_j^{(1)}|\text{EPR}\rangle^{\otimes 2}}{\langle\text{EPR}|^{\otimes 2}U_R(\alpha) |\text{EPR}\rangle^{\otimes 2}}=\frac{\cos{\alpha}}{2}.
  \end{equation}
  Here, we have used the relation $U_R(\alpha)\chi_j^{(1)}=(\cos \alpha \chi_j^{(1)}-\sin \alpha \chi_j^{(2)})U_R(\alpha)$. Applying this procedure to all components yields the expression for $G^{0,\alpha}_{ss'}$ given in the main text. The kernel of the quadratic term in \eqref{eq:act} is therefore the inverse of this free Green's function.

  The remaining steps follow the standard SYK approach~\cite{maldacena2016remarks}. We first integrate out the random couplings,  yielding $\mathcal{Z}(\alpha,t)=\int  \prod_{j=1}^ND\chi_{j}(s)~e^{-S'}$ with action
  \begin{equation}
  \begin{aligned}
  S'=&\int dsds'~\sum_{j}\frac{1}{4}\chi_{j}(s)(G^{0,\alpha})^{-1}\chi_{j}(s')+\frac{J^2}{8N^3}\int dsds'~\left(\frac{1}{2}\sum_j\chi_j(s)\chi_j(s')\right)^4.
  \end{aligned}
  \end{equation}
  Next, we introduce the bilocal field $G_{ss'}=\frac{1}{2N}\sum_j\chi_j(s)\chi_j(s')$ and enforce this relation through a delta function $\delta(G_{ss'}-\frac{1}{2N}\sum_j\chi_j(s)\chi_j(s'))$. We then represent the delta function as a Fourier integral, introducing the self-energy field $\Sigma_{ss'}$. The partition function becomes a functional integral over the Majorana and bilocal fields:
\begin{equation}
\begin{aligned}
\mathcal{Z}(\alpha,t)&=\int D\Sigma_{ss'}DG_{ss'}D\chi_j\,\exp(-S_{\text{eff}}[G,\Sigma]),\\
S_{\text{eff}}&=\int dsds'\,\sum_j\frac{1}{4}\chi_j(s)[(G^{0,\alpha})^{-1}-\Sigma_{ss'}]\chi_j(s')
+\frac{N}{2}\int dsds'\,\left(\Sigma_{ss'}G_{ss'}-\frac{J^2}{4}G_{ss'}^4\right).
\end{aligned}
\end{equation}
Finally, integrating out the Majorana fermions yields the effective $G-\Sigma$ action, up to a field-independent normalization constant:
\begin{equation}
\begin{aligned}
S_{\text{eff}}[G,\Sigma]
=-\frac{N}{2}\operatorname{Tr}\ln\left[(G^{0,\alpha})^{-1}-\Sigma\right]+\frac{N}{2}\int dsds'\,\left(\Sigma_{ss'}G_{ss'}-\frac{J^2}{4}G_{ss'}^4\right).
\end{aligned}
\end{equation}
Here, $\operatorname{Tr}$ denotes the trace over the contour coordinates. We have omitted an $\alpha$-independent constant, which can be fixed by the normalization condition $\mathcal{Z}(0,t)=1$. Varying this action with respect to $G$ and $\Sigma$ gives the saddle-point equations presented in the main text. Once the saddle-point solutions are obtained, we evaluate the partition function using the on-shell action. The expression in the main text follows by applying the saddle-point equations to both terms.

\section{Experimental protocols for measuring the replica twist entropy}
\label{sec:measurement}

In this section, we describe two experimental protocols for measuring the replica twist entropy of a pure state $|\psi\rangle$ with fixed fermion parity. The first uses randomized measurements on repeated preparations of a single copy, while the second uses occupation measurements on two identical copies. For convenience, we denote the replica twist overlap by
\begin{equation}
\label{eq:measurement_overlap}
F(\alpha,\psi)\equiv e^{-\mathcal{P}(\alpha,\psi)}
=\langle\psi|^{\otimes 2}U_R(\alpha)|\psi\rangle^{\otimes 2}.
\end{equation}
Both protocols first determine $F(\alpha,\psi)$ and subsequently obtain the replica twist entropy by taking its negative logarithm.

\emph{Randomized measurements.--}
The first protocol follows from the expansion of the replica twist overlap in Majorana strings. Introducing the total squared correlation strength at each even operator size,
\begin{equation}
\label{eq:measurement_strings}
A_{2k}=\sum_{|S|=2k}
\left|\langle\psi|\chi_S|\psi\rangle\right|^2,
\end{equation}
we write
\begin{equation}
\label{eq:measurement_expansion}
F(\alpha,\psi)=\sum_{k=0}^{N}
\cos\left(\frac{\alpha}{2}\right)^{2N-2k}
\sin\left(\frac{\alpha}{2}\right)^{2k}A_{2k}.
\end{equation}
The quantities $A_{2k}$ can be extracted using randomized Gaussian measurements, employing the measurement ensemble introduced for fermionic classical shadows~\cite{Zhao2021FermionicShadows}.

Specifically, we sample a Gaussian unitary $V$ whose action on the Majorana operators is a uniformly random signed permutation with determinant $+1$. We apply $V$ to $|\psi\rangle$ and measure the occupations of the $N$ complex fermions $a_\ell=(\chi_{2\ell-1}+i\chi_{2\ell})/2$, obtaining a bit string $\bm{b}=(b_1,\ldots,b_N)$. Repeating the preparation and measurement with the same $V$ gives an independent bit string $\bm{b}'$. For each pair of outcomes, we define
\begin{equation}
\label{eq:measurement_parities}
R_k(\bm{b},\bm{b}')
=\frac{1}{\binom{N}{k}}
\sum_{\substack{I\subseteq\{1,\ldots,N\}\\|I|=k}}
\prod_{\ell\in I}(-1)^{b_\ell+b_\ell'}.
\end{equation}
At fixed $V$, the product of parity outcomes from the two independent measurements estimates the square of the corresponding correlation function. Averaging over the random signed permutations samples all Majorana strings with size $2k$ uniformly. Consequently,
\begin{equation}
\label{eq:measurement_average}
\mathbb{E}_{V,\bm{b},\bm{b}'}
[R_k(\bm{b},\bm{b}')]
=\frac{A_{2k}}{\binom{2N}{2k}}.
\end{equation}
Substituting this identity into Eq.~\eqref{eq:measurement_expansion}, we obtain
\begin{equation}
\label{eq:measurement_randomized}
F(\alpha,\psi)=
\mathbb{E}_{V,\bm{b},\bm{b}'}
\left[
\sum_{k=0}^{N}\binom{2N}{2k}
\cos\left(\frac{\alpha}{2}\right)^{2N-2k}
\sin\left(\frac{\alpha}{2}\right)^{2k}
R_k(\bm{b},\bm{b}')
\right].
\end{equation}
The experimental estimator is obtained by replacing this expectation value with an average over measurement records. 
\vspace{10pt}

\emph{Two-copy occupation measurements.--}
The second protocol directly measures the generator of replica rotations. We prepare two identical copies of $|\psi\rangle$ and introduce complex fermions formed by pairing corresponding Majoranas between the replicas:
\begin{equation}
\label{eq:measurement_replica_modes}
f_j=\frac{\chi_j^{(1)}+i\chi_j^{(2)}}{2},
\qquad
n_j=f_j^\dagger f_j
=\frac{1+i\chi_j^{(1)}\chi_j^{(2)}}{2},
\qquad j=1,\ldots,2N.
\end{equation}
Here, Majorana operators in different replicas anticommute. Defining the total occupation $K=\sum_{j=1}^{2N}n_j$, we find
\begin{equation}
\label{eq:measurement_generator}
\Lambda=-\frac{i}{2}\sum_{j=1}^{2N}
\chi_j^{(1)}\chi_j^{(2)}=N-K,
\qquad
U_R(\alpha)=e^{i\alpha(N-K)}.
\end{equation}
The replica twist overlap is therefore the characteristic function of the occupation distribution:
\begin{equation}
\label{eq:measurement_counting}
F(\alpha,\psi)
=\sum_{K=0}^{2N}p(K)e^{i\alpha(N-K)}
=\sum_{K=0}^{2N}p(K)\cos[\alpha(N-K)].
\end{equation}
Here, $p(K)$ is the probability of measuring total occupation $K$, and the second equality follows from the reality of the overlap. Experimentally, we measure the commuting pair parities $i\chi_j^{(1)}\chi_j^{(2)}$. Recording the total occupation $K_m$ in each of $M$ independent repetitions gives the estimator
\begin{equation}
\label{eq:measurement_direct}
\widehat{F}(\alpha,\psi)
=\frac{1}{M}\sum_{m=1}^{M}
\cos[\alpha(N-K_m)].
\end{equation}
Thus, a single occupation distribution determines the entire angle dependence. We emphasize that the protocol requires the distribution of occupation outcomes.

\bibliography{suppl.bbl}